\documentclass[11pt]{article}
\usepackage{fullpage,amsmath}
\usepackage[square,numbers]{natbib}
\usepackage{url}
\usepackage{amssymb, amsmath,natbib,bm,graphicx}
\usepackage{setspace}
\usepackage[plain,noend]{algorithm2e}
\usepackage{multirow}
\usepackage{graphicx}
\usepackage{float}
\usepackage{verbatim}
\usepackage{color}
\usepackage{xcolor}
\usepackage{mathtools}

\usepackage[colorlinks,urlcolor=blue,linkcolor = blue]{hyperref}
\usepackage{currfile-abspath}

\getmainfile % get real main file (can be different than jobname in some cases)
\getabspath{\themainfile} % or use \jobname.tex instead (not as safe)
\usepackage{booktabs}
\usepackage{bigstrut}
\title{Unifying Boxplots: A Multiple Testing Perspective}
\author{
	Bowen Gang\\
	\small Department of Statistics and Data Science, Fudan University, Shanghai, China\\
	\small \texttt{bgang@fudan.edu.cn}
	\and
	Hongmei Lin\\
	\small School of Statistics and Data Science, Shanghai University of International Business and Economics\\
	\small \texttt{hmlin@suibe.edu.cn}
	\and
	Tiejun Tong\\
	\small Department of Mathematics, Hong Kong Baptist University, Hong Kong, China\\
	\small \texttt{tongt@hkbu.edu.hk}
}
\date{} % Suppress date

\begin{document}
	
	\maketitle
	
	\begin{abstract}
		Tukey's boxplot is a foundational tool for exploratory data analysis, but its classic outlier-flagging rule does not account for the sample size, and subsequent modifications have often been presented as separate, heuristic adjustments. In this paper, we propose a unifying framework that recasts the boxplot and its variants as graphical implementations of multiple testing procedures. We demonstrate that Tukey's original method is equivalent to an unadjusted procedure, while existing sample-size-aware modifications correspond to controlling the Family-Wise Error Rate (FWER) or the Per-Family Error Rate (PFER). This perspective not only systematizes existing methods but also naturally leads to new, more adaptive constructions. We introduce a boxplot motivated by the False Discovery Rate (FDR), and show how our framework provides a flexible pipeline for integrating state-of-the-art \textcolor{black}{robust estimation} techniques directly into the boxplot's graphical format. By connecting a classic graphical tool to the principles of multiple testing, our work provides a principled language for comparing, critiquing, and extending outlier detection rules for modern exploratory analysis.
	\end{abstract}
	

\section{Introduction}

%This template helps you to create a properly formatted \LaTeXe\ manuscript.
%%%%%%%%%%%%%%%%%%%%%%%%%%%%%%%%%%%%%%%%%%%%%%
%% `\ ' is used here because TeX ignores    %%
%% spaces after text commands.              %%
%%%%%%%%%%%%%%%%%%%%%%%%%%%%%%%%%%%%%%%%%%%%%%
Tukey's box-and-whisker plot remains a cornerstone of exploratory data analysis celebrated for its elegant distillation of a sample's key features \citep{tukey1977exploratory}. Its outlier flagging rule, however, based on a fixed multiple of the interquartile range, does not account for the sample size.\footnote{\textcolor{black}{It is worth noting that the interpretation of the boxplot's fences as a definitive rule for identifying "outliers" is a later development. Tukey's original proposal was more nuanced, distinguishing between "outside" observations (those between 1.5 and 3 interquartile ranges (IQRs) from the quartiles) and "far out" observations (more than 3 IQRs away), without explicitly labeling either as outliers. In its modern application, both "outside" and "far out" points are typically flagged simply as potential outliers.}} This property, while simple, leads to a procedural artifact where the number of "outliers" detected in samples from a pure normal distribution grows with the number of observations, a behavior inconsistent with modern inferential standards.

Over the decades, numerous authors have proposed modifications to make the boxplot’s outlier-detection mechanism adaptive to the sample size \citep{schwertman2007identifying,banerjee2007simple,barbato2011features,barbato2009outlier,hofmann2017value}. \textcolor{black}{While these proposals vary in their approach, with some replacing the fence rule entirely, many are introduced as heuristic adjustments.} For this latter group, a closer examination reveals a notable, though typically implicit, relationship to the core principles of multiple hypothesis testing. 
For example, procedures that control the "some-outside rate per sample" the probability of incorrectly flagging at least one \textcolor{black}{non-outlying observation} as an outlier are conceptually equivalent to controlling the Family-Wise Error Rate (FWER) \citep{sim2005outlier,hoaglin1986performance,schwertman2004simple}. More recently, the "Chauvenet-type boxplot" introduced by \cite{lin2025tukey} implicitly aligns with the control of the Per-Family Error Rate (PFER), as it aims to limit the expected number of false positives to a small constant.

In this paper, we argue that these are not isolated parallels but rather different manifestations of a single, unifying idea: the boxplot can be viewed as a graphical implementation of a multiple testing procedure. This perspective provides a powerful and coherent framework for understanding, comparing, and extending boxplot methodologies. The classic Tukey boxplot represents an unadjusted procedure, while its successors correspond to different, more stringent modes of error control.
The primary contribution of this work is the establishment of this unifying framework. Once the boxplot is seen through this lens, its potential is dramatically expanded. As a first natural consequence, we demonstrate how to construct a boxplot motivated by the False Discovery Rate (FDR), the canonical error metric for modern large-scale inference \citep{benjamini1995controlling}. This yields fences that adapt not only to the sample size but also to the data themselves.
Perhaps more importantly, this conceptual bridge connects a classic graphical tool to the rich and evolving literature on multiple testing. It transforms the boxplot from a rigid set of outlier-detection rules into a flexible and adaptive inferential framework. By incorporating recent advances in robust estimation and applying $p$-value-based multiple testing procedures, researchers can readily tailor the boxplot to their specific analytical needs, enhancing both its relevance and effectiveness in modern data analysis.

\section{A Multiple Testing Framework for Boxplots: A Conceptual Review}
The core observation of this short paper is that the evolution of the boxplot's outlier-detection rules can be systematically understood as the application of different multiple testing error-control standards. 
\textcolor{black}{To formalize this, let us consider a sample of observations $\{X_1, \dots, X_n\}$. To motivate our framework, we assume these are independent and identically distributed (iid) draws from a continuous distribution. This allows us to conceptually model the sample as a potential mixture: the bulk of the data drawn from one distribution, with a small fraction of outliers arising from one or more contaminating distributions.}

For each observation, we can define a null hypothesis, $H_{0i}$, that states \textcolor{black}{"$X_i$ is not an outlier"}, meaning it is drawn from the same underlying distribution as the bulk of the data. The act of flagging $X_i$ as a potential outlier is then equivalent to rejecting $H_{0i}$. With $n$ such tests being performed simultaneously, the central question becomes how to manage the rate of Type I errors. Let $V$ be the number of \textcolor{black}{non-outlying observations} that are incorrectly flagged as outliers, i.e., the number of true null hypotheses that are rejected. Let also $Q_1$ and $Q_3$ be the 1st and 3rd quartiles of the samples,\footnote{\textcolor{black}{There are multiple ways to define sample quartiles. Throughout this paper, we use the sample quartile definition corresponding to Definition 7 in \cite{hyndman1996sample}, as it is the default in the R function \texttt{quantile} and widely adopted in practice.}} respectively, and $\text{IQR}=Q_3-Q_1$ be the interquartile range.
We can now map existing boxplot variations to specific strategies for controlling $V$.

\subsection{The Unadjusted Procedure: Tukey's Classic Boxplot}
Tukey's original boxplot defines its lower and upper {inner} fences at $\text{LF}=Q_1 - 1.5 \times \text{IQR}$ and $\text{UF}=Q_3 + 1.5 \times \text{IQR}$, respectively. For data from a normal distribution, the probability of an observation falling outside these two fences is approximately 0.7\%. In the language of hypothesis testing, this corresponds to setting a Per-Comparison Error Rate (PCER)---the probability of a Type I error for a single test---at a fixed level of $0.007$. 

Critically, this procedure makes no adjustment for multiplicity. The threshold is constant regardless of the number of tests, $n$. Consequently, the expected number of falsely flagged outliers, $E(V)$, is approximately $0.007n $, which grows linearly with the sample size. This lack of adjustment is the defining feature of an unadjusted multiple testing procedure and explains the well-known tendency of the classic boxplot to flag an excessive number of points in large datasets.

\subsection{Controlling the FWER}

The most stringent form of multiple testing control is the management of the FWER, defined as the probability of making at least one Type I error:
\[
\text{FWER} = P(V \geq 1).
\]
This is precisely the conceptual goal of boxplot modifications that seek to control the "some-outside rate per sample" \citep{hoaglin1986performance, sim2005outlier}. These methods derive a sample-size-dependent fence coefficient, $k_n$, to define the outlier region $\mathcal{O}_n = (-\infty, \mathrm{LF}_n) \cup (\mathrm{UF}_n, \infty)$, where
\begin{equation*}
	\mathrm{LF}_n = Q_1 - k_n \times \text{IQR} \quad \text{and} \quad \mathrm{UF}_n = Q_3 + k_n \times \text{IQR}.
\end{equation*}
Considerable effort has been dedicated to numerically solving for the value of $k_n$ that ensures
\begin{equation*}
	P( \text{at least one of } X_1, \dots, X_n \in \mathcal{O}_n ) \le \alpha,
\end{equation*}
for some given $\alpha$, assuming \textcolor{black}{the bulk of the data} are drawn from a normal distribution \citep{hoaglin1987fine, sim2005outlier}. 
%While conceptually sound, this approach requires bespoke numerical solutions for a given distributional assumption.

\subsection{Controlling the PFER}
A less conservative Type I error notion is the PFER, which controls the expected number of Type I errors:
\[
\text{PFER} = E(V).
\]
The Chauvenet-type boxplot recently proposed by \cite{lin2025tukey} is a direct implementation of this principle. The method's fence coefficient, $k_n$, is derived from Chauvenet's criterion, which historically identifies an observation as an outlier if its $p$-value is less than $0.5/n$. For $n$ tests, this corresponds to controlling the PFER at a level of $n \times (0.5/n) = 0.5$. The fence coefficient is set to
\[
k_n = \frac{\Phi^{-1}(1 - 0.25/n)}{1.35}-0.5,
\]
where $\Phi^{-1}$ is the quantile function of the standard normal distribution, and the constants 0.5 and 1.35 arise from using quartiles to estimate the mean and standard deviation. 
\textcolor{black}{By construction, assuming the non-outlying data are normally distributed, this method aims to produce, on average, half a false positive per dataset, regardless of the sample size.}

%By construction, this method aims to produce, on average, half a false positive per dataset, regardless of sample size.

\section{A Unified $p$-Value Pipeline for Boxplot Construction}\label{sec3}

The conceptual framework above, which connects boxplots to multiple testing, naturally leads to a unified and flexible methodology for the fence construction. Instead of deriving fence coefficients $k_n$ for each error metric, we propose a general pipeline that transforms the outlier detection problem into a standard $p$-value-based multiple testing problem. This approach is not only conceptually simpler but also vastly more extensible. The pipeline consists of four steps as follows.

\begin{enumerate}
	\item \textbf{Parameter Estimation.} 
	First, we obtain robust estimates of the parameters governing the distribution of the \textcolor{black}{non-outlying data}. Assuming the bulk of the data follows a normal distribution, we can robustly estimate its location $\mu$ and scale $\sigma$ using the sample quartiles as recommended by \cite{tukey1977exploratory,higgins2019cochrane}\footnote{In the case where $Q_1 = Q_3$, which implies $\text{IQR} = 0$, our estimator $\hat{\sigma}$ becomes zero. This typically occurs with discrete or heavily rounded data. When this happens, we recommend using a scale estimator based on the Median Absolute Deviation (MAD) to ensure a non-zero denominator whenever the IQR is zero but the data are not all identical \cite{pham2001mean}. That is,
		$
		\hat{\sigma}=\text{MAD}/0.675,
		\quad \text{where}\ 
		\text{MAD}=\text{median}(|X_i-\text{median}(X_1,\ldots,X_n)|).
		$
	}:
	\begin{equation}\label{eq:est}
		\hat{\mu} = \frac{Q_1+Q_3}{2} \quad \text{and} \quad \hat{\sigma} = \frac{Q_3-Q_1}{1.35}.
	\end{equation}
	
	\item \textbf{$p$-value Calculation.} Next, we convert each observation $X_i$ into a two-sided $p$-value based on the estimated null distribution:
	$$
	p_i = 2 \left( 1 - \Phi\left(\left|\frac{X_i - \hat{\mu}}{\hat{\sigma}}\right|\right) \right).
	$$
	This transforms the raw data into a set of $p$-values $\{p_1, \dots, p_n\}$, which serve as standardized evidence against the null hypothesis of \textcolor{black}{not being an outlier}.
	
	\item \textbf{Multiple Testing Adjustment.} The set of $n$ $p$-values is fed into any standard $p$-value-based multiple testing procedure to control a desired error rate (e.g., FWER, PFER, FDR) at a level $\alpha$. The procedure returns a significance threshold, $t_{\text{adj}}$, which is potentially data-dependent. Any hypothesis $H_{0i}$ with $p_i \le t_{\text{adj}}$ is rejected by the chosen testing procedure.
	
	\item \textbf{Fence Construction.} Finally, to translate this decision rule back into the graphical language of a boxplot, we determine the $z$-score corresponding to the $p$-value threshold, $z_{\text{adj}} = \Phi^{-1}(1 - t_{\text{adj}}/2)$. The fences are then defined directly on the original data scale:
	\begin{align*}
		\mathrm{LF}_n &= \hat{\mu} - z_{\text{adj}} \cdot \hat{\sigma}=Q_1-(z_{\text{adj}}/1.35-0.5)\times\text{IQR}, \\
		\mathrm{UF}_n &= \hat{\mu} + z_{\text{adj}} \cdot \hat{\sigma}=Q_3+(z_{\text{adj}}/1.35-0.5)\times\text{IQR}.
	\end{align*}
\end{enumerate}

\textcolor{black}{If prior knowledge suggests that outliers are likely to appear only on one side of the distribution, the procedure can be adapted by using one-sided $p$-values. For example, to detect large outliers, the $p$-value calculation in Step 2 would be modified to $p_i = 1 - \Phi((X_i - \hat{\mu})/\hat{\sigma})$. Consequently, the fence construction in Step 4 becomes asymmetric. The $z$-score threshold is now $z_{\text{adj}} = \Phi^{-1}(1 - t_{\text{adj}})$, which defines an upper fence at $\mathrm{UF}_n = \hat{\mu} + z_{\text{adj}} \cdot \hat{\sigma}$. No lower fence is statistically defined by this test, so the lower whisker simply extends to the sample minimum. }

This pipeline provides a principled way to construct boxplots. The choice of error metric and control procedure in Step 3 directly determines the behavior of the resulting outlier-flagging rule, allowing for transparent design and easy extension.

%\section{Instantiating the Framework: FWER, PFER, and FDR Boxplots}
%Our unified pipeline can be used to construct boxplots that align with various multiple testing philosophies, from classic conservative approaches to modern adaptive ones.

\subsection{The FWER Boxplot}
To create a boxplot that controls the FWER, we simply apply a standard FWER-controlling procedure in Step 3 of our pipeline. For instance, using the Bonferroni or the more powerful Holm method \cite{holm1979simple} on the calculated $p$-values at level $\alpha$ will yield a threshold $t_{\text{FWER}}$. The resulting fences will graphically represent the rejection region of the chosen FWER procedure. This approach elegantly circumvents the need for complex numerical simulations to find $k_n$ and connects the boxplot directly to established inferential machinery.

\subsection{The PFER Boxplot}
The principle behind the Chauvenet-type boxplot can also be implemented through our pipeline. To control the PFER at a level $\gamma$ (e.g., $\gamma=0.5$), one would reject any $p$-value smaller than $t_{\text{PFER}} = \gamma/n$. This is equivalent to taking the $z$-value threshold $z_{\text{adj}}$ in Step 4 of the general pipeline to be $\Phi^{-1}(1 - \gamma/(2n))$. If $\mu$ and $\sigma$ are estimated using Eq.~\eqref{eq:est}, we recover the  Chauvenet-type boxplot as in \cite{lin2025tukey}.

\subsection{A Natural Extension: The FDR Boxplot}
Having established this flexible pipeline, a natural and powerful extension is to construct a boxplot that controls the False Discovery Rate (FDR), the canonical error metric for modern large-scale inference \citep{benjamini1995controlling}. The FDR is the expected proportion of false discoveries among all rejected hypotheses. Let $R$ be the total number of observations flagged as outliers and $V$ be the number of \textcolor{black}{non-outlying observations} that are incorrectly flagged as such. The FDR is defined as
\[
\text{FDR} = E\left[\frac{V}{R}\right], \quad \text{with } \frac{V}{R} \equiv 0 \text{ if } R=0.
\]
To construct an FDR-controlling boxplot, we apply an FDR procedure, such as the Benjamini-Hochberg (BH) method, in Step 3 of our pipeline. This yields a data-dependent threshold $t_{\text{FDR}}$. The fences are then set based on this threshold:
\begin{align*}
	\mathrm{LF}_n &= \hat{\mu} - z_{\text{\tiny FDR}} \cdot \hat{\sigma}=Q_1-(z_{\text{\tiny FDR}}/1.35-0.5)\times\text{IQR} \\
	\mathrm{UF}_n &= \hat{\mu} + z_{\text{\tiny FDR}} \cdot \hat{\sigma}=Q_3+(z_{\text{\tiny FDR}}/1.35-0.5)\times\text{IQR},
\end{align*}
where $z_{\text{\tiny FDR}} = \Phi^{-1}(1 - t_{\text{\tiny FDR}}/2)$.

The key advantage of the FDR boxplot is its \textbf{adaptivity}. If the data contain many true outliers with very small $p$-values, the BH procedure will yield a larger threshold $t_{\text{FDR}}$. This results in narrower fences and thus greater power to detect additional outliers. Conversely, if the data contain few or no outliers, the threshold will be small, resulting in wider, more conservative fences. This dynamic behavior makes the FDR boxplot a uniquely powerful tool for modern exploratory analysis, and its straightforward derivation showcases the practical utility of our unifying framework.

\subsection{A Toy Example: From Data to Fences}
To make this pipeline concrete, consider a small dataset of $n=11$ observations as
$$ X = \{9, 16,  18, 20, 20, 22, 22, 24,26, 36, 50\}. $$
\textcolor{black}{In this dataset, the observation 50 is well-separated from the bulk of the data, while 9 and 36 can be regarded as borderline cases.}
We will walk through the pipeline to see how the  FWER, PFER, and FDR controls yield different flagged outliers and significance threshold $t_{\text{adj}}$ with the control levels being $\alpha=0.01$ for FWER/FDR and $\gamma=0.5$ for PFER. \\

\noindent\textbf{Steps 1-2: Parameter Estimation and $p$-value Calculation.}
\begin{itemize}
	\item The sample quartiles are $Q_1=19$ and $Q_3=25$, giving $\text{IQR} = 6$.
	\item From Eq.~\eqref{eq:est}, we have $\hat{\mu} = (19+25)/2 = 22$ and $\hat{\sigma} = 6/1.35 \approx 4.44$.
	\item The eleven $p$-values, sorted in ascending order, are:
	\begin{align*}
		p_{(1)} &\approx 2.98\times 10^{-10} \text{ (for 50)},\\
		p_{(2)}& \approx 1.63\times 10^{-3} \text{ (for 36)},\\
		p_{(3)} &\approx 3.44\times 10^{-3} \text{ (for 9)},\\
		p_{(4)}& \approx 0.177 \text{ (for 16)}, \dots 
	\end{align*}
	The rest of the $p$-values are all greater than 0.36.
\end{itemize}

\noindent\textbf{Steps 3-4: Multiple Testing and Fence Construction.}
Now we apply different adjustment procedures to find the significance threshold $t_{\text{adj}}$.
\begin{itemize}
	\item \textbf{PFER (Chauvenet):} The threshold is fixed: $t_{\text{PFER}} = \gamma/n = 0.5/11 = 0.045$. The first three $p$-values ($p_{(1)}, p_{(2)}, p_{(3)}$) are all well below this threshold. This flags $\{50, 36, 9\}$ as outliers. 
	%The significance threshold is $0.05$.
	\item \textbf{FDR (BH) at $\alpha=0.01$:} The BH procedure finds the largest $i$ such that $p_{(i)} \le i\alpha/n = i(0.01)/11 $.
	We have
	$p_{(1)} \le 1(0.01)/11$,
	$p_{(2)} \le 2(0.01)/11$, and
	$p_{(3)} \approx 0.00344 \not\le 3(0.01)/11 $.
	That is, the largest $i$ is 2. The procedure rejects the first two hypotheses, flagging $\{50, 36\}$, and the significance threshold is $1.63\times 10^{-3}$.
	\item \textbf{FWER (Holm) at $\alpha=0.01$:} The Holm procedure compares $p_{(i)}$ to $\alpha/(n-i+1) = 0.01/(12-i)$.
	We have
	$p_{(1)} < 0.01/11$ and
	$p_{(2)} \not\le0.01/10$. It thus rejects only the first hypothesis, flagging just $\{50\}$ as outlier. The significance threshold is $2.98\times 10^{-10}$.
\end{itemize}

\noindent\textbf{Summary of Results.}
This toy example clearly illustrates the distinct behaviors of boxplots under different error-control philosophies. In this example with small $n$, the PFER approach is the most liberal. The FDR method, even at a strict $\alpha=0.01$, is adaptive enough to detect two outliers due to the strong evidence from the first. The FWER method, being the most conservative, stops short and only flags 50 as an outlier at this strict $\alpha$ level. The resulting fences, shown in Table~\ref{tab:toy_example}, are starkly different. For reference, Tukey's classic method, which relies on a fixed 1.5$\times$IQR rule (corresponding to $t_{\text{adj}}=0.007$) rather than a statistical adjustment, would have flagged $\{50,36,9\}$ in this instance.

\begin{table*}[t] 
	\centering
	\caption{Comparison of significance threshold and outliers flagged for the toy example.}
	\label{tab:toy_example}
	\begin{tabular}{@{}lccc@{}}
		\hline
		Method &  $t_{\text{adj}}$ & Outliers Flagged & Lower/Upper Fences \\ \hline
		PCER (Tukey) & $0.007$ & $\{50, 36, 9\}$ & $[10, 34]$ \\
		PFER (Chauvenet) at 0.5 & $0.05$ & $\{50, 36, 9\}$ & $[13.3, 30.7]$ \\
		FDR (BH) at 0.01 & $p_{(2)} \approx 0.00163$ & $\{50, 36\}$ & $[8, 36]$ \\
		FWER (Holm) at 0.01 & $p_{(1)} \approx 2.98\times 10^{-10}$ & $\{50\}$ & $[-6, 50]$ \\ \hline
	\end{tabular}
\end{table*}

\section{Numerical Experiments}

To visually demonstrate the distinct behaviors of different outlier detection rules, we conduct simulation studies that compare five types of boxplots. We compare the unadjusted Tukey boxplot, the three boxplots derived from our framework (controlling FWER, PFER, and FDR), and the method proposed by \cite{barbato2009outlier} (hereafter, BGL).

\subsection{Implementation Details}\label{sec:4.1}
The five methods are implemented as follows. The FWER, PFER, and FDR methods all rely on the same $p$-value generation pipeline, differing only in the final multiple testing adjustment.

\begin{itemize}
	\item\textbf{PCER-type (Tukey):} The classic boxplot with fences defined at $Q_1 - 1.5 \times \text{IQR}$ and $Q_3 + 1.5 \times \text{IQR}$.
	\item\textbf{FWER-type (Holm):} The Holm procedure \citep{holm1979simple}  is used for testing at a target FWER level of $\alpha = 0.01$.
	\item\textbf{PFER-type (Chauvenet):} We use the Chauvenet-type boxplot from \cite{lin2025tukey}, which is constructed to control the PFER at a target level of $0.5$.
	\item\textbf{FDR-type (BH):} The BH procedure \citep{benjamini1995controlling} is used for testing at a target FDR level of $\alpha = 0.01$.
	\item\textbf{BGL-type:} \textcolor{black}{The method from \cite{barbato2009outlier}, which makes the fences sample-size-dependent. The two fences are defined at $Q_1 - 1.5 \times \text{IQR} \times [1 + 0.1 \log(n/10)]$ and $Q_3 + 1.5 \times \text{IQR} \times [1 + 0.1 \log(n/10)]$.}
\end{itemize}
For the three $p$-value-based methods (Holm, Chauvenet, and BH), the $p$-value for each observation is calculated relative to an assumed normal distribution. The parameters of this distribution are defined by the robust quartile-based estimators from Eq.~\eqref{eq:est}.
\textcolor{black}{The implementations of the FWER-type (Holm) and FDR-type (BH) boxplots are available in our newly developed R package \texttt{AdaptiveBoxplot}, which can be found on GitHub at \url{https://github.com/bgang92/AdaptiveBoxplot}.}
\subsection{\textcolor{black}{Normally Distributed Majority}}
We generate data from a normal mixture model to simulate a common scenario: \textcolor{black}{the bulk of the observations} are drawn from a standard normal $N(0,1)$ distribution, but 1\% are contaminating "true outliers" from a $N(5,1)$ distribution. The formal model is as follows:
\begin{align*}
	\theta_i &\overset{iid}{\sim} \text{Bernoulli}(0.01),\\
	X_i|\theta_i &\overset{ind}{\sim} (1-\theta_i)N(0,1) + \theta_i N(5,1),
\end{align*}
where $i=1,\ldots, n$.
We simulate datasets with sample sizes $n=5\times 10^k$ for $k=1,2,3$. 
%The results are visualized in Figure~\ref{fig:normal_mixture}.

\begin{figure*}[ht!]
	\centering
	\includegraphics[width=\linewidth,height=5in]{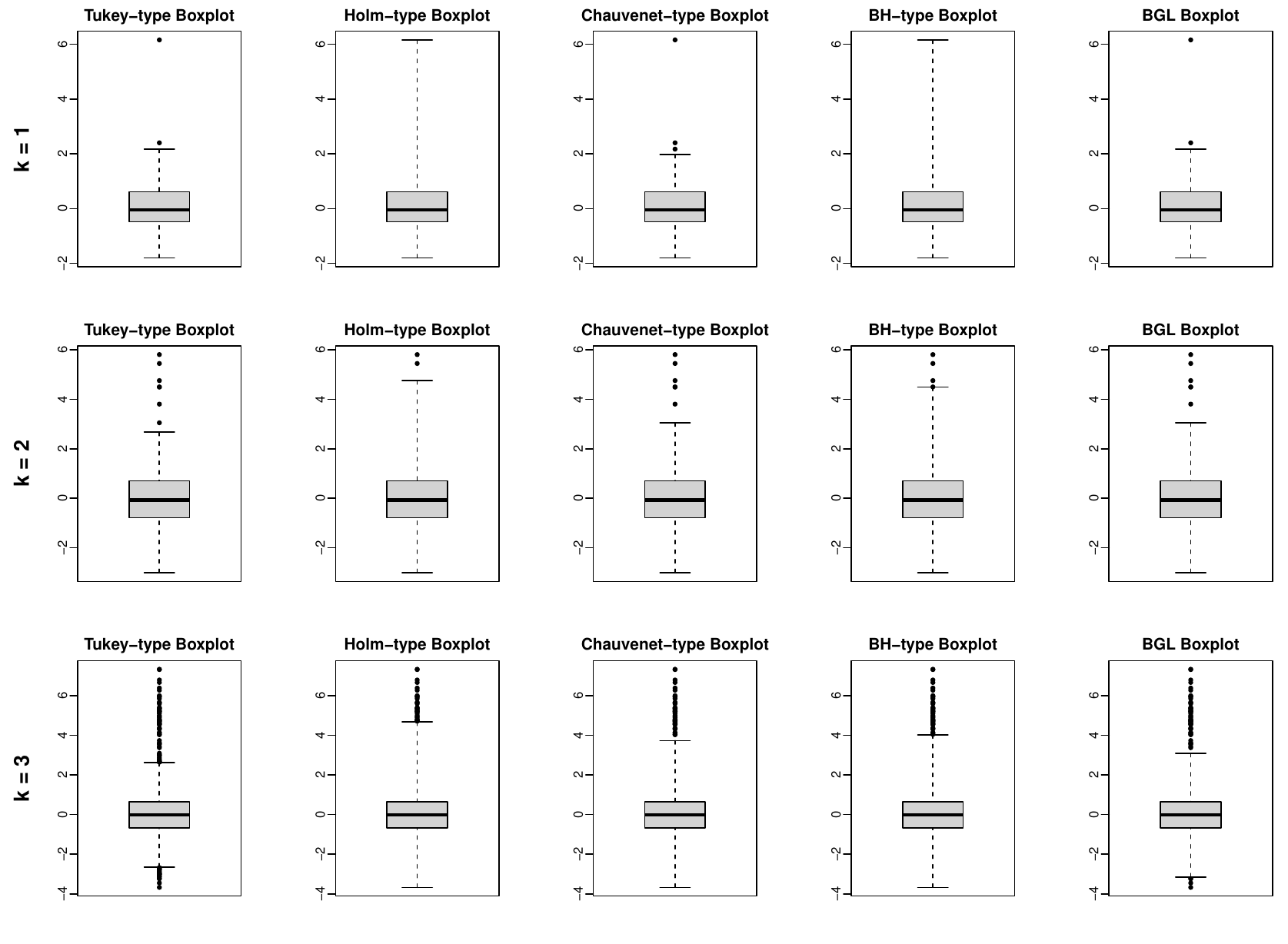} % Assuming the image file is named normal.pdf
	\caption{A visual comparison of the five boxplot methodologies across different sample sizes. The data are from a normal mixture model with 1\% true outliers. The behavior of each boxplot aligns perfectly with its underlying error control principle.}
	\label{fig:normal_mixture}
\end{figure*}
Figure~\ref{fig:normal_mixture} provides a visual confirmation of our framework. The failure of the Tukey-type boxplot at large sample sizes is evident. As $n$ increases from 50 to 5000, 
the number of observations flagged from the $N(0,1)$ distribution explodes, flooding the plot with false positives and obscuring the true outliers.
In contrast, all four adjusted methods adapt their fences to account for the sample size, preventing the flood of false discoveries seen with the fixed coefficient approach.

While the figure gives a qualitative overview, a deeper analysis of the fence coefficients in Table~\ref{tab:normal} reveals the core mechanistic differences between the methods. This table highlights two distinct levels of adaptivity.
\textcolor{black}{
	The Chauvenet (PFER) and BGL boxplots illustrate the first level: their fence coefficients are deterministic functions of the sample size $n$. Both become more conservative as $n$ grows, a clear improvement over the fixed Tukey fences, but they remain blind to the actual content of the data.}

In contrast, the FWER and FDR boxplots exhibit a more profound, data-driven adaptivity. Their fence coefficients are not fixed by $n$ alone but are determined by the empirical distribution of the $p$-values from the sample. This is powerfully illustrated by the fence coefficient for FDR (BH) in Table~\ref{tab:normal}. As $n$ increases from 50 ($k=1$) to 500 ($k=2$), the coefficient increases to guard against false positives. However, as $n$ grows to 5000 ($k=3$), the coefficient surprisingly decreases. This is the signature of FDR's adaptivity: at $n=5000$, there are approximately 50 strong outlier signals. The BH procedure detects this abundance of outliers, becomes more powerful, and sets a more liberal threshold (a smaller $z$-score), resulting in tighter fences to capture additional, less extreme outliers. The Holm-FWER method, bound by its goal of preventing even a single error, cannot leverage this information as efficiently as BH-FDR.

This analysis shows that the choice is not merely between fixed and adaptive fences, but between different philosophies of adaptation: one that is pre-determined by the sample size, and another that dynamically responds to the evidence within the data itself.

\begin{table}[t] 
	\centering
	\caption{Comparison of fence coefficients for data from a normal mixture model. The coefficients are averaged over 5000 simulation replicates.}
	\label{tab:normal}
	\begin{tabular}{@{}lccc@{}}
		\hline
		Method &  $k=1$ &$k=2$ &$k=3$ \\ \hline
		PFER (Chauvenet) at 0.5 & $1.41$ & $1.93$ & $2.38$ \\
		FDR (BH) at 0.01 & $2.08$ & $2.82$ & $2.46$ \\
		FWER (Holm) at 0.01 & $2.11$ & $3.02$ & $3.06$ \\ 
		BGL & $1.60$ & $1.75$ & $1.90$ \\ \hline
	\end{tabular}
	
\end{table}

\subsection{Performance on Skewed Data}\label{sec:chisq}
We now consider a scenario where the data are drawn from a single, skewed distribution.  We generate $n$ observations directly from a chi-square distribution with 10 degrees of freedom, $\chi^2_{10}$. This distribution is right-skewed, meaning that even without true mixture-model outliers, there is a heavy upper tail that can be taken for outliers.

\begin{figure*}[ht!]
	\centering
	\includegraphics[width=\linewidth,height=4.5in]{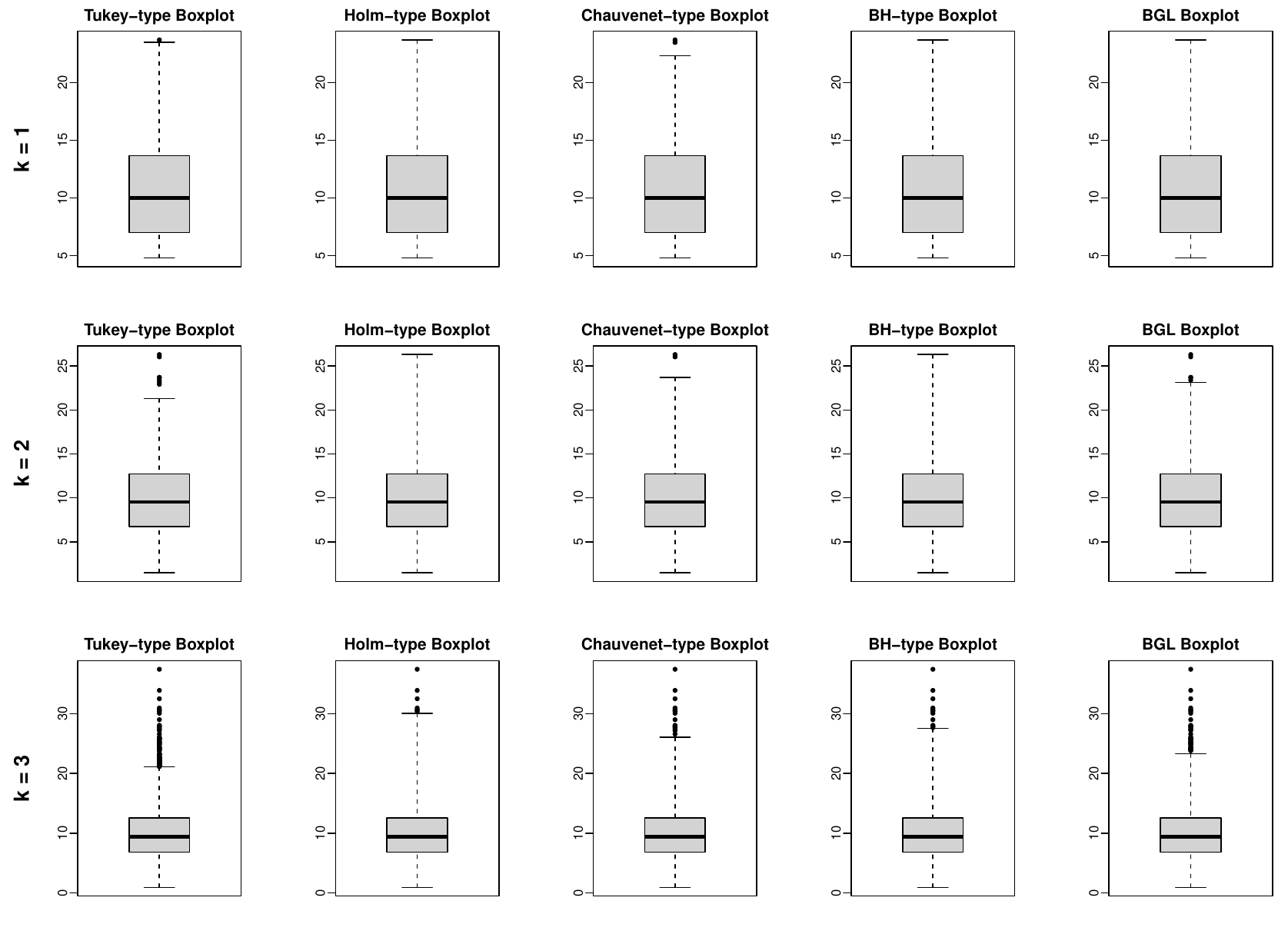} 
	\caption{Performance of the five boxplot methodologies on right-skewed data generated from a $\chi^2_{10}$ distribution.}
	\label{fig:chisq}
\end{figure*}

Crucially, for the FWER, PFER, and FDR-type boxplots, we deliberately maintain the simple, normal-based procedure from the previous section. That is, we continue to estimate the location and scale parameters using Eq.~\eqref{eq:est} and proceed as if the majority of the data were sampled from a normal distribution. This allows us to investigate the practical performance of the methods under model misspecification. We again simulate datasets with sample sizes $n=5 \times 10^k$ for $k=1, 2, 3$. The results are summarized in Figure~\ref{fig:chisq} and Table~\ref{tab:chisq}.

As the sample size $n$ increases, all five methods begin to flag a growing number of points in the upper tail as outliers. This is the expected outcome when a symmetric model is imposed on an asymmetric distribution. The right skewness ensures that many genuine, albeit extreme, observations from the $\chi^2_{10}$ distribution will fall far above the symmetrically-placed upper fence.
We observe a similar hierarchy of conservatism as in the normal case. The {Tukey-type boxplot}, being unadjusted, flags the most points. The {Holm-type boxplot} is the most conservative, flagging the fewest points due to its stringent FWER control, which results in the widest fences. The {Chauvenet-type}, {BH-type}, and BGL boxplots lie in between.

Table~\ref{tab:chisq} quantifies this and reveals deeper insights when compared to the results from the normal case (Table~\ref{tab:normal}). The PFER and BGL fence coefficients are identical in both tables, this is not surprising given the two methods are blind to the data's distribution and depend only on the sample size. In contrast, the FDR and FWER methods are clearly responsive. The FDR boxplot, in particular, is more adaptive than the FWER boxplot.
It interprets the flood of small $p$-values from the heavy tail as evidence of abundant "outliers", causing it to become more liberal. Consequently, its fence coefficient only increases slightly when the sample size increases from 500 to 5000. This demonstrates that the FDR procedure is working as designed.

This simulation reveals the diagnostic power of our framework. When a normal-based procedure is applied to skewed data, all methods are naturally affected. However, the problem is not a flaw in the multiple testing logic itself, but rather in the initial model of the data used to generate $p$-values. The challenge of improving the boxplot for non-normal data is thus cleanly reframed as a more familiar statistical task: selecting a more appropriate probability distribution for the main body of the data. The adjustment machinery is sound; its inputs must simply be more accurate.

\begin{table}[t] 
	\centering
	\caption{Comparison of fence coefficients for $\chi^2_{10}$ data, fence coefficients computed by averaging results over 5000 simulation replicates.}
	\label{tab:chisq}
	\begin{tabular}{@{}lccc@{}}
		\hline
		Method &  $k=1$ &$k=2$ &$k=3$ \\ \hline
		PFER (Chauvenet) at 0.5 & $1.41$ & $1.93$ & $2.38$ \\
		FDR (BH) at 0.01 & $1.82$ & $2.65$ & $2.78$ \\
		FWER (Holm) at 0.01 & $1.84$ & $2.73$ & $3.17$ \\ 
		BGL & $1.60$ & $1.75$ & $1.90$ \\ \hline
	\end{tabular}
	
\end{table}

\textcolor{black}{\section{Generalizing the Framework: Beyond Normality}}\label{sec:5}
\textcolor{black}{So far our analysis has proceeded under the working assumption that the bulk of the data is drawn from a normal distribution. This is not an arbitrary choice, but rather a robust and pragmatic starting point for exploratory analysis. Its justification is rooted in Winsor's principle, which famously states that "\textit{All distributions are normal in the middle}" \cite{tukey1960survey}. 
	However, the power of the $p$-value pipeline lies in its flexibility, not its adherence to a single distribution. If prior knowledge suggests that the bulk of the data is better described by a different parametric family, which we denote generally as $F(\cdot; \theta)$, the framework can and should be adapted. Here, $F$ is the cumulative distribution function and $\theta$ is the vector of its parameters (e.g., location, scale, shape). The procedure is generalized as follows:
	\begin{enumerate}
		\item \textbf{Parameter Estimation:} We obtain robust estimates 
		of the parameters of the chosen parametric family, denoted as $\hat{\theta}$.
		\item \textbf{$p$-value Calculation:} We compute $p$-value for each observation using the estimated distribution function, $F(X_i; \hat{\theta})$. 
		%This step allows for important refinements; for instance, when modeling with a distribution known to be skewed to the right, a one sided $p$-value (e.g., $p_i = 1 - F(X_i; \hat{\theta})$) may be more appropriate than a two-sided $p$-value.
		\item \textbf{Multiple Testing Adjustment:} With the new set of $p$-values, we apply the chosen multiple testing procedure (e.g., FWER, PFER, FDR) to obtain an adjusted significance threshold, $t_{\text{adj}}$. Any hypothesis for which $p_i \le t_{\text{adj}}$ is rejected.
		\item \textbf{Fence Construction:} Finally, the decision rule is translated back into the graphical language of the boxplot. The fences are constructed directly from the quantiles of the fitted distribution $F(\cdot,\hat{\theta})$.
		When two-sided $p$-value is used, the rejection region is split between both tails:
		\begin{align*}
			&\mathrm{LF}_n=F^{-1}(t_{\text{adj}}/2; \hat{\theta}),\\
			& \mathrm{UF}_n= F^{-1}(1 - t_{\text{adj}}/2; \hat{\theta}).
		\end{align*}
\end{enumerate}}
\textcolor{black}{As noted in Section \ref{sec3}, this pipeline can be made more appropriate for skewed data by using one-sided $p$-values. In the general case, the $p$-value calculation in Step 2 is modified to test for outliers in a specific tail. For a right-sided test (to detect large outliers), the $p$-value is $p_i = 1 - F(X_i; \hat{\theta})$. Consequently, the fence construction in Step 4 is asymmetric, defining only an upper fence at $\mathrm{UF}_n = F^{-1}(1 - t_{\text{adj}}; \hat{\theta})$. For a left-sided test (to detect small outliers), the $p$-value is $p_i = F(X_i; \hat{\theta})$, which defines only a lower fence at $\mathrm{LF}_n = F^{-1}(t_{\text{adj}}; \hat{\theta})$. In either one-sided case, the whisker on the non-tested side simply extends to the sample minimum or maximum. }

\textcolor{black}{To demonstrate the practical utility of this generalized framework, we now conduct a simulation study. We use the same data generating process from Section \ref{sec:chisq}, where datasets are drawn from a $\chi^2_{10}$ distribution and contain no contaminating outliers. An ideal outlier detection procedure should therefore flag very few, if any, observations. We compare the performance of the five methods mentioned in Section \ref{sec:4.1}. }

\begin{figure*}[ht!]
	\centering
	\includegraphics[width=\linewidth,height=4.5in]{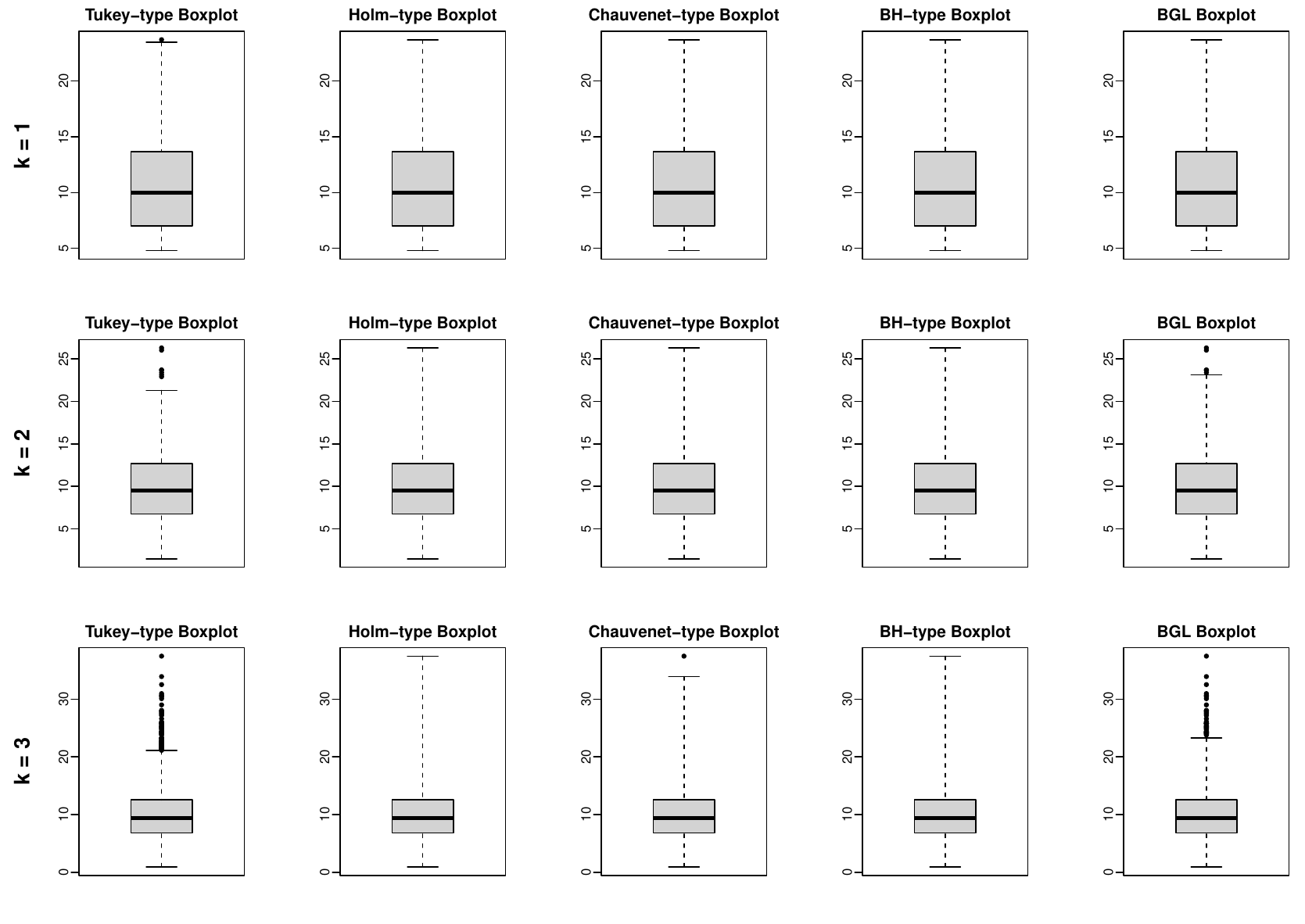} 
	\caption{Performance of the five boxplot methodologies on right-skewed data generated from a $\chi^2_{10}$ distribution, assuming the model is correctly specified.}
	\label{fig:chisqcorrect}
\end{figure*}

\textcolor{black}{To estimate the degrees of freedom parameter, we use a robust estimator based on the highly accurate Wilson Hilferty approximation for the median of a $\chi^2$ variable \cite{wilson1931distribution}, which states that $\text{median}(\chi^2_k) \approx k(1-2/(9k))^3$. Accordingly, $k$ is estimated by numerically solving the equation
	\[
	\text{median}(X_1,\ldots, X_n) = \hat{k}\left(1 - \frac{2}{9\hat{k}}\right)^3.
	\]
	To compute $p$-values, since the $\chi^2_k$ distribution is skewed to the right, we use the one-sided $p$-value
	\[
	p_i = 1 - F(X_i; \hat{k}),
	\]
	where $F(\cdot,\hat{k})$ is the distribution function of the $\chi^2_{\hat{k}}$ random variable. 
	These $p$-values are then used to construct two boxplots: one using the Holm procedure to control the FWER and another using the BH procedure to control the FDR, both at a target level of 0.01. 
	For the Chauvenet-type boxplot, we define the upper fence at $F^{-1}(1-0.5/n,\hat{k})$ and the lower fence at $\min\{X_1,\ldots,X_n\}$.
	The results are summarized in Figure~\ref{fig:chisqcorrect}. }

\textcolor{black}{
	The figure provides a striking and unambiguous confirmation of our generalized framework's utility. The Tukey and BGL boxplots, which are both based on an assumption of symmetry, consequently flag a substantial number of observations in the upper tail. This effect becomes more pronounced with increasing sample size, leading to a visualization containing numerous false positives that can obscure the true structure of the data.}
\textcolor{black}{In stark contrast, the methods based on the correctly specified $\chi^2$ model perform exceptionally well. The Holm-type and BH-type boxplots achieve near-perfect performance, flagging no observations as outliers. The Chauvenet-type boxplot is also highly effective, flagging only a small number of the most extreme observations. All three methods demonstrate a vast improvement over the normal-based approaches because they use an appropriate reference distribution, correctly recognizing that the heavy upper tail is an intrinsic feature of the data. }
%They therefore avoid the flood of false positives seen with the other methods and produce a clean plot that accurately reflects the underlying distribution.

\section{Concluding Remarks}

This paper reframes the boxplot, transforming it from a collection of disparate, heuristic rules into a coherent and extensible statistical methodology. For decades, the evolution of the boxplot has been characterized by ad-hoc, sample-size-specific adjustments. We have shown that many seemingly isolated modifications can be understood and improved through the unified lens of multiple hypothesis testing. Our primary contribution is the development of a general $p$-value pipeline that operationalizes this insight. 
This pipeline reveals a powerful underlying unity: it shows that all major boxplot variations can be generated by a single fence formula $(z_{\text{adj}}/1.35-0.5)\times\text{IQR}$, whose sensitivity is governed solely by an effective $z$-score, $z_{\text{adj}}$. Remarkably, Tukey's classic $1.5 \times \text{IQR}$ rule can be regarded as a special case, corresponding to a fixed $z_{\text{adj}} = 2.7$. Our pipeline, therefore, not only systematizes existing methods by mapping them to explicit error control principles (PCER, FWER, PFER) but, more importantly, provides an engine for creating new, more powerful diagnostic tools. The power of this unified methodology is demonstrated through its immediate practical payoffs. We introduced the FDR boxplot, a novel construction that brings the canonical error metric of modern large-scale science to a classic exploratory data analysis tool.

The practical utility of this framework lies in its modularity and flexibility. As we demonstrated in Section \ref{sec:5}, the pipeline is not limited to the assumption of normality but provides a flexible scaffold for building principled, context-aware outlier detection tools for any parametric family. This transforms the boxplot from a static summary graphic into a dynamic diagnostic tool, where its performance offers direct insight into the appropriateness of the underlying distributional assumptions.

\textcolor{black}{Further refinements to our $p$-value pipeline are readily available by incorporating more advanced estimation techniques in Step 1. While our analysis relied on simple, robust quartile-based estimators, the framework can be significantly enhanced by employing methods from the rich literature on robust statistics. This field provides a wealth of tools designed to yield reliable parameter estimates in the presence of data contamination. For comprehensive treatments of robust methods, from foundational theory to modern applications, see, for example, \cite{hampel1986robust,HuberRonchetti2009,maronna2019robust}.
	In the specific but common case where the bulk of the data is assumed to follow a normal distribution, a powerful alternative is the empirical null methodology. This data-adaptive approach, developed for large-scale inference, provides a suite of tools for accurately estimating the parameters of the majority normal distribution directly from the data \citep{efron2004large, jin2007estimating, gauran2018empirical}.}

Despite the power of these refinements, it is important to contextualize the role of the boxplots discussed here.
They remain, at their heart, tools for {exploratory data analysis}. 
While their designs are motivated by and aligned with formal notions of Type I error, we do not claim that they rigorously control these error rates in a formal inferential sense. 
The procedures we have outlined rely on robust but simple estimates of an {assumed distribution for the bulk of the data}, and true statistical control would require a more formal handling of parameter estimation uncertainty and potential dependencies among observations. Such challenges are best addressed in subsequent, more formal modeling stages rather than at the exploratory phase.

Acknowledging this important distinction, the conceptual bridge we have drawn between graphical diagnostics and multiple testing remains powerful. It provides a principled language for discussing, comparing, and critiquing different outlier detection rules. More broadly, this framework offers a powerful lens for innovation that extends in two key directions. First, its principles can be applied to other graphical diagnostics; many plots, from Q-Q plots to residual plots, involve multiple visual comparisons that could be enhanced with ideas from adaptive error control. Second, while this paper has focused on univariate data, the core pipeline is highly general. We believe it offers a promising path for creating sample-size-aware boxplots for more complex data structures, such as functional data \citep{hyndman2010rainbow,sun2011functional,dai2018functional,qu2022sparse}, circular data \citep{abuzaid2012boxplot, buttarazzi2018boxplot}, and curve data or paths \citep{mirzargar2014curve,raj2017path}. Exploring these avenues is a rich direction for future research and lies beyond the scope of this paper.
\bibliographystyle{chicago} % Style BST file (imsart-number.bst or imsart-nameyear.bst)
\bibliography{refs.bib}       % Bibliography file (usually '*.bib')

%% or include bibliography directly:

\end{document}